\documentclass[aps,prb,superscriptaddress,twocolumn]{revtex4-2}
\usepackage{graphics}
\usepackage{epsfig}
\usepackage{color}
\usepackage{stackrel}
\usepackage{amsfonts}
\usepackage{wrapfig}
\usepackage{pstricks}
\usepackage{pst-node}
\usepackage{bm}
\usepackage{dcolumn}
\usepackage{multirow}
\usepackage{epstopdf}
\usepackage{subfigure}
\usepackage{amssymb}
\usepackage{amsmath}
\usepackage{commath}
\usepackage{graphicx,bm}
\usepackage{verbatim}
\usepackage{booktabs}
\usepackage[para]{threeparttable}
\usepackage{etoolbox}
\usepackage{lipsum}

\begin{document}
\title{Tetrons, pexcitons, and hexcitons in monolayer transition-metal dichalcogenides}

\author{Dinh~Van~Tuan}
\email[]{vdinh@ur.rochester.edu}
\affiliation{Department of Electrical and Computer Engineering, University of Rochester, Rochester, New York 14627, USA}
\author{Hanan~Dery}
\altaffiliation{hanan.dery@rochester.edu}
\affiliation{Department of Electrical and Computer Engineering, University of Rochester, Rochester, New York 14627, USA}
\affiliation{Department of Physics and Astronomy, University of Rochester, Rochester, New York 14627, USA}

\begin{abstract}
We present a comprehensive theoretical analysis of composite excitonic states in doped transition-metal dichalcogenide monolayers. Making use of the pair distribution function, we introduce a method to include the effect of screening in the interaction between the excitonic state and Fermi-sea electrons. Employing the screened potential, we study  tetrons and pexcitons in which a trion is bound to one and two holes in the conduction band, respectively. A conduction-band hole denotes the lack of electrons with certain quantum numbers at the vicinity of the trion. We then analyze the hexciton complex wherein conduction-band  holes facilitate binding of three electrons with one valence-band hole. We introduce  a simple model from which one can readily calculate the binding energy of the third (satellite) electron in the hexciton complex. Finally, we compare the simulated results with experiment and point out the shortcomings and successes of the analysis.
\end{abstract}

\maketitle

\section{Introduction} \label{sec:intro} 

A valence-band hole is causing a stir when introduced to an electron-doped semiconductor through photoexcitation. When the electron density is relatively small, a conduction-band (CB) electron can bind to the valence-band (VB) hole and its photoexcited electron to create a charged exciton (trion) \cite{Kheng_PRL93,Finkelstein_PRL95}. When the electron density is relatively large, Mahan showed that the interaction between an infinite-mass  VB hole and Fermi-surface electrons  leads to singularity in the optical conductivity \cite{Mahan_PR67a,Mahan_PR67b,Skolnick_PRL87}.  The singularity is muted if the mass of the VB hole is not much larger than the mass of electrons  \cite{Hawrylak_PRB91}. The evolution from three-body trions at vanishing electron densities to muted Fermi-edge singularities at large densities goes through a regime at which Fermi-sea electrons and the trion become correlated \cite{Bronold_PRB00,Suris_PSS01,Esser_pssb01}.

Transition-metal dichalcogenide monolayers are an ideal platform to study the evolution of correlated trions \cite{Efimkin_PRB17,Chang_PRB18,Chang_PRB19,Rana_PRB20}. Compared with typical semiconductors, electron-hole (\textit{e-h}) pairs are strongly bound in these monolayers due to relatively large effective masses of CB electrons and VB holes as well as moderate dielectric constants. Combined with the impeded dielectric screening in two-dimensional (2D) systems, the binding energy  between an exciton and resident CB electron is of the order of few tens meV \cite{Xu_NatPhys14,Wang_RMP18}. As a result, the evolution of trions can be studied across a relatively wide range of  charge densities, starting from zero up to a few times 10$^{12}$~cm$^{-2}$. At low temperatures, there is a density regime in these monolayer semiconductors in which the trion binding energy is larger than the Fermi energy, which in turns exceeds the thermal energy ($\varepsilon_T > E_F \gg k_BT$). Namely, the Fermi sea is past its incipient stage.

In the regime that $E_F \gg k_BT$, a CB electron can bind to the VB hole by vacating a state below the Fermi surface and sample a relatively large portion of the $k$-space above it. The electron can then repeatedly interact with the VB hole and stay close to it. Just as promoting an electron from the VB to the CB during photoexcitation leaves behind an unfilled state in the VB (hole), pulling out an electron from below the Fermi surface leaves behind a hole in the Fermi sea. This CB hole stays close to the pulled out electron by scattering between occupied $k$-states below the Fermi surface.  The binding between  photoexcited \textit{e-h} pair and the CB electron-hole pair (\textit{e-}$\bar{e}$) is possible if the electrons are quantum-mechanically distinguishable. The two electrons can then stay together near the VB hole without violating the Pauli exclusion principle. In electron-doped GaAs, for example, photoexcited electrons belong to one of two reservoirs of resident CB electrons: spin-up and spin-down states around the $\Gamma$-point of the Brillouin zone. As a result, the  photoexcited \textit{e-h} pair strongly interacts with a CB pair whose electron has opposite spin  (i.e., \textit{e$_\uparrow$-h} with \textit{e$_\downarrow$-$\bar{e}$} or \textit{e$_\downarrow$-h} with \textit{e$_\uparrow$-$\bar{e}$}). 

Generalizing this concept, composite excitonic states can emerge if the photoexcited pair interacts simultaneously with \textit{multiple} distinguishable electrons \cite{s,l}. Made possible by their accompanying CB holes, distinguishable electrons can stay together near the VB hole at the same time.  In this work, we present a comprehensive theoretical analysis of 4-, 5-, and 6-body composites (tetrons, pexcitons and hexcitons) in electron-doped transition-metal dichalcogenide monolayers. We present a method to include the effect of screening in the interaction between particles of an excitonic complex and CB holes. The calculation of excitonic states is treated with the screened potential and the results are compared with the case that screening is ignored.  In addition, we introduce a simple model from which one can readily calculate the binding energy when a satellite electron joins a correlated trion to form a hexciton. 

This work is part of a tetrad. Of the three studies that accompany this paper, one is analysis of experimental results in ML-WSe$_2$ \cite{h}. Using magneto-reflectance data, we show that trions evolve to 6-body (hexcitons) and then to 8-body composite states (oxcitons) when the photoexcited \textit{e-h} pair  interacts with three distinguishable CB pairs.  In addition, we explain how the central and secondary optical transitions of hexcitons manifest in the photoluminescence data.  The evidence we provide in that study weakens our previous argument that exciton interaction with shortwave plasmons stands behind the observed  optical transitions in electron rich ML-WSe$_2$  \cite{Dery_PRB16,VanTuan_PRX17,Scharf_JPCM19,VanTuan_PRB19}.  The second study is a short description of the theory of composite excitonic states in doped semiconductors \cite{s}. We present central pieces of the theoretical framework, explain how to include CB holes in the calculation of composite excitonic states, and highlight important results of the theory. The third study elaborates on the theoretical formulation and  computational details of the stochastic variational method in momentum space (SVM-{\it k}), with which we calculate composite excitonic states \cite{l}. In addition to the direct Coulomb interaction between quasiparticles of the composite, the SVM-{\it k} allows us to incorporate the band-gap renormalization effect and electron-hole exchange interaction of the  \textit{e-}$\bar{e}$ pairs. 

The organization of this paper is as follows. Section~\ref{sec:trion} includes analysis of  a negative trion whose two electrons are restricted to momentum space above the Fermi surface. We examine how Pauli blocking of low energy states  affects  the binding energy and wavefunction of the trion. Section~\ref{sec:screening}  deals with the way at which screening is incorporated in the interaction between the trion and  Fermi-sea electrons. The screened potential is employed in Sec.~\ref{sec:tetron_pexciton} to study correlated trions (tetrons and pexcitons), made of trions and CB holes. Section~\ref{sec:hex} includes analysis of hexcitons wherein CB holes facilitate binding of three electrons with one VB hole. In addition,  we present a simple model in which a satellite electron binds to a correlated trion. Section \ref{sec:conc} provides outlook and conclusions. We explain what are the shortcomings of the analysis and what future works should address.  The Appendix includes a compiled list of parameters used in this work and in Refs.~\cite{s,h}.

\section{Bare trions} \label{sec:trion} 

A bare trion is a three-particle complex consisting of a resident charge carrier from the semiconductor that is bound to the photoexcited \textit{e-h} pair. Solutions of this three-body problem can explain various observations in the absorption and emission spectra of monolayer transition-metal dichalcogenides. For example,  the calculated binding energy of the trion with respect to the exciton matches the energy difference between the corresponding resonances of their optical transitions \cite{Kezerashvili_FBS17,Mayers_PRB15,Kylanpaa_PRB15,Kidd_PRB16,Donck_PRB17,Mostaani_PRB17,VanTuan_PRB18}. In addition, the fine structure  of negative trions in WSe$_2$ monolayers can be explained through the exchange interaction between the CB electron and VB hole \cite{Hichri_PRB20,Glazov_JCP20}. Finally, the difference in binding energies of positive and negative trions in WSe$_2$ monolayers can be explained through the different mass of the two electrons in optically-active negative trions \cite{VanTuan_PRB18}. 

Ideally, the bare trion picture is most accurate when the photoexcited \textit{e-h} pair is generated in a semiconductor with one CB electron or one VB hole. Yet, experiments show that when the charge density increases, many-body signatures slowly evolve from the trion optical transition. Therefore, the trion is a good starting point to study photoexcitation in doped semiconductors. In this section, we analyze the behavior of a negative trion whose two electrons are restricted to momentum space above the Fermi surface due to Pauli blocking of low energy states.  Interactions between Fermi-sea electrons and the trion's particles are to be introduced in later sections of this work. By omitting these interactions but considering Pauli blocking, we will examine the change to  the binding energy of the trion due to band filling.  

 \begin{figure}
\includegraphics[width=8cm]{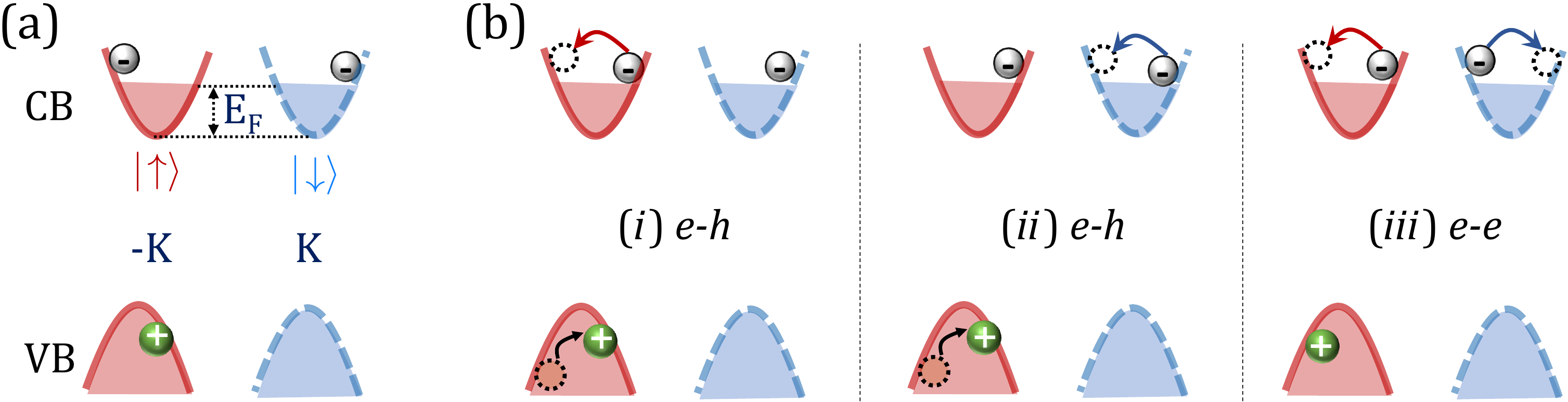}
\caption{Bare trions whose electrons are restricted to $k$-space states above the  Fermi surface ($k > k_F$). (b)  Coulomb processes leading to binding of the three particles in the trion. Interactions between Fermi-sea electrons and the trion's particles through CB holes are analyzed in later sections.} \label{fig:trion}
\end{figure}

\begin{figure*}
\includegraphics[width=16cm]{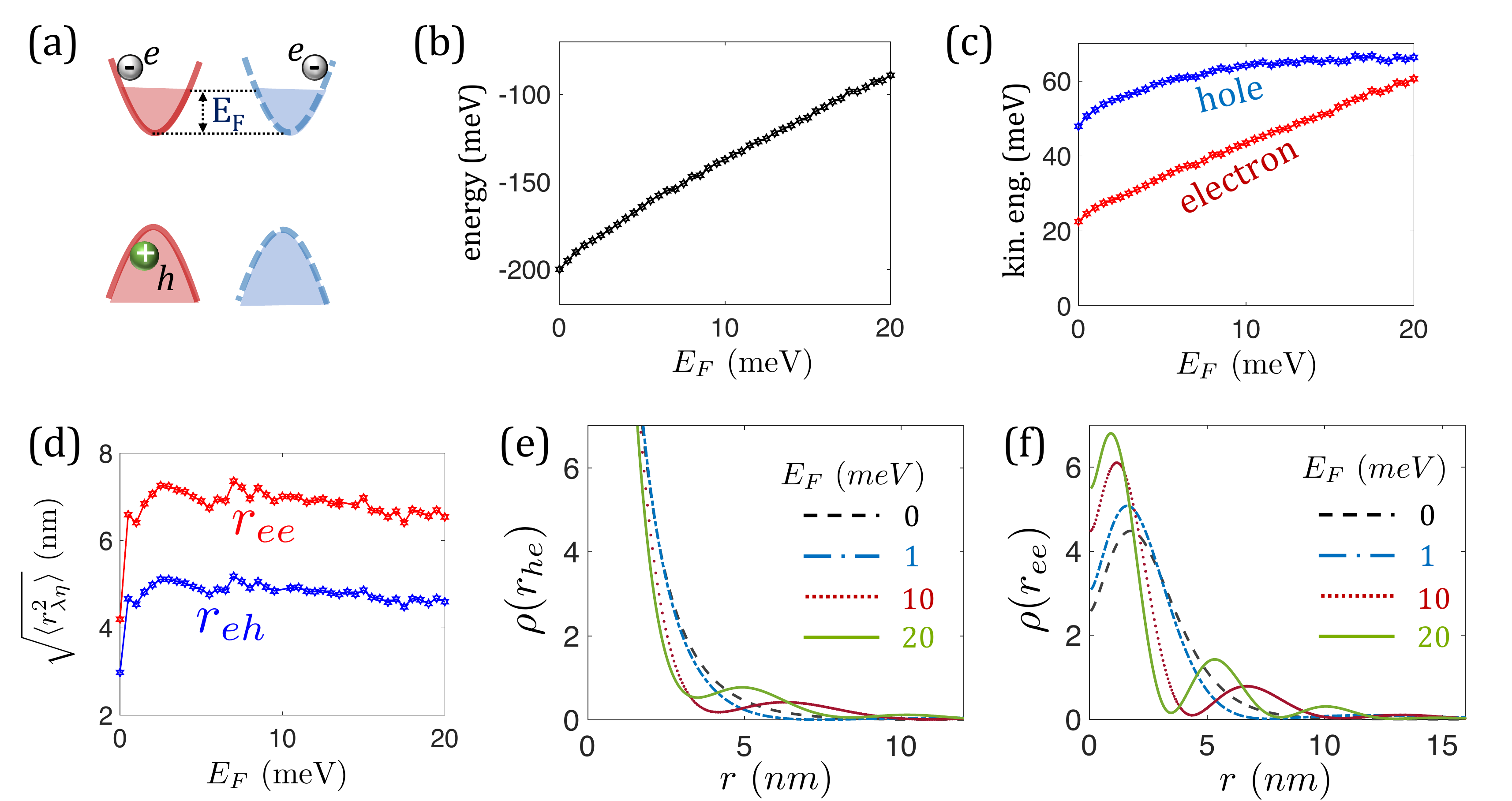}
\caption{(a) A bare trion whose two electrons are restricted in $k$-space above the Fermi surface.  (b) Binding energy vs Fermi energy. (c) Kinetic energies of the hole and each of the two electrons. (d) Average distances between the VB hole and each of the electrons and between the two electrons. (e) and (f) Density distributions of the electron-hole and electron-electron separations, respectively, plotted at four different Fermi energies. } \label{fig:core}
\end{figure*}

Figure~\ref{fig:trion}(a) shows the trion complex we have in mind. Interactions between the three particles of the trion, shown in Fig.~\ref{fig:trion}(b), are assumed to be governed by the bare Coulomb potential. In the context of the Coulomb potential, `bare' means unscreened by other electrons. Use of the bare Coulomb potential is applicable if the gained energy from binding of the three particles is much larger than $E_F$. The relative motion of the three particles in this limit is too fast for Fermi-sea electrons to respond and screen the electric fields between the three particles of the trion. A similar result can be inferred by taking the limit $\omega \rightarrow \infty$ of the dynamically screened potential, which again boils down to the bare Coulomb potential \cite{Scharf_JPCM19}.  Unless otherwise noted,  the bare Coulomb potential  is modeled by the 2D Keldysh-Rytova potential  \cite{Rytova_PMPA67,Keldysh_JETP79,Cudazzo_PRB11},
\begin{equation}
V_b(q) = \frac{2 \pi e_\lambda e_\nu}{A \epsilon(q) q} , \label{eq:KRP}
\end{equation} 
where $A$ is the area of the ML, and $e_{\lambda(\eta)}$ is the charge of particle $\lambda(\eta)$. The dielectric function, 
\begin{equation}
\epsilon(q) = \epsilon_b(1 + r_0 q) = \epsilon_b + r^{\ast} q  ,   \label{eq:eps_q}
\end{equation}
depends on the dielectric constant of the barriers that encapsulate the ML, $\epsilon_b$, and the polarizability of the ML $r_0$ where $r^{\ast}  =  \epsilon_b r_0$.

We use the SVM-$k$ to calculate the behavior of a bare trion whose electrons are restricted to states above the Fermi surface \cite{l}.  The results are summarized in Figs.~\ref{fig:core}(b)-(f).  The binding energy, shown in Fig.~\ref{fig:core}(b),  drops from $\sim$200 to  $\sim$90~meV when the Fermi energy increases from 0 to 20 meV (electron density increases from 0 to $\sim 4\times10^{12}$~cm$^{-2}$).  The energy decays faster than the band-filling effect because it is harder for the VB hole to bind with faster electrons (large $k$). Figure~\ref{fig:core}(c) shows the kinetic energy of the particles, where the electron component refers to the kinetic energy of each of the two electrons (not their sum). The total kinetic energy of the three particles is roughly  doubled when the Fermi energy increases from 0 to 20~meV. Figure~\ref{fig:core}(d) shows the average distances between the particles, from which it is noticeable that the size of the trion is  largely unaffected by increasing the charge density. This apparent contradiction between a non-bloated wavefunction and large drop in binding energy is understood if we plot the density distributions for distances between the particles.  Figures~\ref{fig:core}(e) and (f) show these results, from which we can notice sharpening of the slopes when the charge density increases. The slopes are directly related to the enhanced kinetic energy ($\mathbf{p} \propto \nabla$). 

These results suggest that trions are resilient in the sense that they do not bloat when electrons are added to the semiconductor. Their spatial extents do not differ appreciably compared with those of excitons because the hole is equally and strongly bound to either of the two electrons \cite{Mayers_PRB15,Kylanpaa_PRB15,Kidd_PRB16,Donck_PRB17,Mostaani_PRB17,VanTuan_PRB18}. The trion is glued by short-range forces between its three particles, whereas the long-range dipolar force between an exciton and electron plays a secondary role.  As such, it is misleading to think of a trion as tightly-bound exciton that is loosely held together with a satellite electron. The three particles of the trion remain strongly bound as long as the bare potential is a good description of the Coulomb interaction between its particles. 

Figure~\ref{fig:core}(b) shows that  the binding energy of the bare trion is rapidly lost when the charge density increases.  This behavior is not seen in experiment, wherein the energy shifts of  optical transitions associated with trions are hardly affected when the Fermi energy is of the order of 10~meV \cite{Wang_NanoLett17,Smolenski_PRL19,Wang_PRX20,Liu_PRL20,Liu_NatComm21,Li_NanoLett22}.  To remedy this problem, we have to consider the Coulomb interaction between the trion and  CB holes.

\section{Screening} \label{sec:screening} 
 This section deals with the way screening is introduced to the interaction between the trion and  CB holes. Recalculation of the binding energies due to the added screened interaction will be discussed in the next sections. As we will show, the binding energy is overestimated when the interaction between the trion and  CB holes is treated through the bare Coulomb potential. On the other hand, the binding energy is underestimated when screening is introduced through the random-phase approximation (RPA). Below, we develop a method to mitigate this problem. 

Following the RPA theory of an electron gas,  the static limit of the screened potential reads 
\begin{equation}
V_{s}(q) =  \frac{2\pi e_\lambda e_\nu}{A  }   \frac{ 1  }{\kappa_q + q\epsilon(q)    } ,   \label{eq:RPA}
\end{equation}
where the Thomas-Fermi wavenumber is  
\begin{equation}
\kappa_q = \frac{2m_b^2}{ \hbar^2  } \left[ 1- \sqrt{1- \frac{4 k_F^2}{q^2}} \,\,\,\Theta \left(q - 2 k_F \right)\right].
\end{equation}
$m_b$ is the effective mass of Fermi-sea electrons and $\Theta(x)$ is the Heaviside step  function. The RPA screening theory  is accurate at very large charge densities $r_s  <1$, where $r_s = e^2m_b/\hbar^2 \sqrt{\pi n}$ is the unitless Wigner–Seitz radius in 2D systems. This density regime corresponds to $n > 10^{14}$ cm$^{-2}$ in transition-metal dichalcogenide monolayers, whereas our interest is in much smaller densities (e.g., $n \sim 10^{12}$ cm$^{-2}$).  Corrections to the screened potential, such as the ones originally offered by Hubbard and later by Singwi \textit{et al.} \cite{Hubbard_RSPA58,Singwi_PR68}, rely on sum rules of the homogeneous electron-gas theory \cite{Giuliani_Vignale_Book,Mahan_Book,Ichimaru_RMP82}. The problem with using these potentials in the context of excitonic complexes is that we are interested in the behavior next to the mobile impurity (e.g., the trion), where the electron gas is by definition inhomogeneous. 

To deal with this problem, we recall that electrons with similar quantum numbers to those of the two electrons in the core trion  are pushed away from the complex due to the Pauli exclusion principle. Furthermore, if resident electrons in the monolayer are hosted in two reservoirs and the core trion includes electrons from both, the result  is that charge density fluctuations are suppressed in the charge-depleted area near the core trion. Thus, screening is no longer effective at distances that are commensurate with $1/k_F$ or less from the complex (the average inter-particle distance in the electron gas is about $1/k_F$). The screening can only manifest at large distances from the VB hole. These conditions are naturally met if the screened potential is defined through the ansatz,
\begin{equation}
V_{g}(r) = \left[ 1 - g(r)\right] V_b(r) + g(r) V_{s}(r)\,\,, \label{eq:vg}
\end{equation}
where $V_b$ and $V_s$ are the bare and RPA-screened potentials, respectively. The limits of the weight function are $g(r \rightarrow 0)=0$ and $g(r \rightarrow \infty)=1$, so that the Coulomb interaction is bare near the trion and screened far from it. A natural choice in our case is the sum of pair distribution functions, 
\begin{equation}
g(r) = \frac{1}{2}\Big(g_{\uparrow\uparrow}(r) + g_{\downarrow\downarrow}(r) \Big). \label{eq:g_sum}
\end{equation}
$g_{ss}(r)$ is the joint probability of finding two particles with similar spin at distance $r$. That is, we assume that the electrons of the core trion are at the origin, and $g(r)$ is the probability to find a spin-up or spin-down electron away from the core trion. The simplest way to evaluate the pair distribution function is to consider the antisymmetry of the wavefunction in a noninteracting 2D electron gas at zero-temperature. When the electron gas is not spin polarized, $g_{\uparrow\uparrow}(r)=g_{\downarrow\downarrow}(r)$, one gets \cite{Giuliani_Vignale_Book}
\begin{equation}
g(r) = 1 - \left| \frac{1}{nA} \sum_{k<k_F} \exp(-i\mathbf{k}\cdot\mathbf{r})\right|^2 = 1 - 4  \frac{J_1^2(k_F r)}{k^2_F r^2}, \label{eq:g}
\end{equation}  
where $n$ is the electron density ($2\pi n = k_F^2$), and $J_1(x)$ is the first-order Bessel function of the first kind. The sum rule
\begin{equation}
\int_0^\infty d^2r [1 - g(r)]n = 2, \label{eq:sum}
\end{equation}  
denotes the fact that we have two electrons at the center of the trion.   The Fourier transform of $g(r)$ reads 
\begin{eqnarray} 
 g({\bf q}) &=& \delta_{\bf q,0}  - \frac{8\Theta (2k_F -q)}{A k^2_F } \times \nonumber \\ &&\,\,\,\,\, \left(  \arccos\left(\frac{q}{2 k_F}\right)- \frac{q}{2 k_F} \sqrt{1-\frac{q^2}{4k_F^2}} \right) . \label{eq:gq}
 \end{eqnarray}
While this function vanishes for $q > 2k_F$, it does not affect the interaction with CB holes: the transferred momentum has to be smaller than $2k_F$ for a CB hole to stay below the Fermi surface.  The Fourier transform of the partially-screened potential in Eq.~(\ref{eq:vg}) has the form
\begin{eqnarray}
V_{g}({\bf q}) &=& V_b({\bf q})  - \sum_{\bf p}  g({\bf p}) \left[ V_b({\bf q- p}) -  V_{s}({\bf q- p})  \right]   \nonumber \\ &\simeq& \frac{2 \pi e_\lambda e_\nu}{A \epsilon(q)} \frac{1}{\sqrt{q^2 + q_0^2}} - \frac{1}{A}\sum_{\bf p}  g({\bf p})  \times  \nonumber \\ && \,\,\,\,\,\left[   \frac{2 \pi e_\lambda e_\nu}{\epsilon(\mathbf{q}+\mathbf{p})} \frac{1}{\sqrt{|\mathbf{q}+\mathbf{p}|^2 + q_0^2}}  -  \frac{\pi \hbar^2}{m_b}  \right] . \label{eq:vg_numer}
\end{eqnarray}
Note the added parameter $q_0$ in the bare-potential  expressions. Its value is finite ($q_0 \neq 0$)  in the presence of disorder where $1/q_0$ is the characteristic distance between localization centers or charge puddles. As we will see, CB holes or satellite electrons are strongly bound to the complex already at very small electron densities in ideal MLs ($q_0 = 0$). The reason is that most of their binding comes from the limit  $q \rightarrow 0$ in which the bare potential diverges. The use of $q_0 \neq 0$ means that the binding of CB holes (or satellite electrons) is mitigated when $k_F < q_0$ because disorder and fluctuations can readily break the correlation between the core trion and the incipient Fermi sea. Disorder is less relevant when it comes to the bare trion because its radius  is much smaller than the characteristic distance between localization centers. 

Quantitatively, the parameter $q_0$ is introduced by adding a decay to the long-range part of the Keldysh-Rytova potential in real space,
\begin{equation}
V(r) = \frac{\pi e^2}{2r^\ast} \left[\mathbf{H}_0\left(\frac{r}{r_0}\right) - Y_0\left(\frac{r}{r_0}\right)\right] \exp(-q_0r). \label{eq:KRP_2} 
\end{equation}
$\mathbf{H}_0(r)$ and $Y_0(r)$ are the zeroth-order Struve function and Bessel function of the second kind, respectively. Assuming that the polarizability radius is much smaller than the characteristic distance between localization centers, $q_0r_0 \ll 1$, we can replace the long-range part of $\mathbf{H}_0(x) - Y_0(x)$ with  $\sim 2/\pi x$. The result is $1/\sqrt{q^2 + q_0^2}$ instead of $1/q$ in the 2D Fourier transform of the potential.

\begin{figure}
\includegraphics[width=8.5cm]{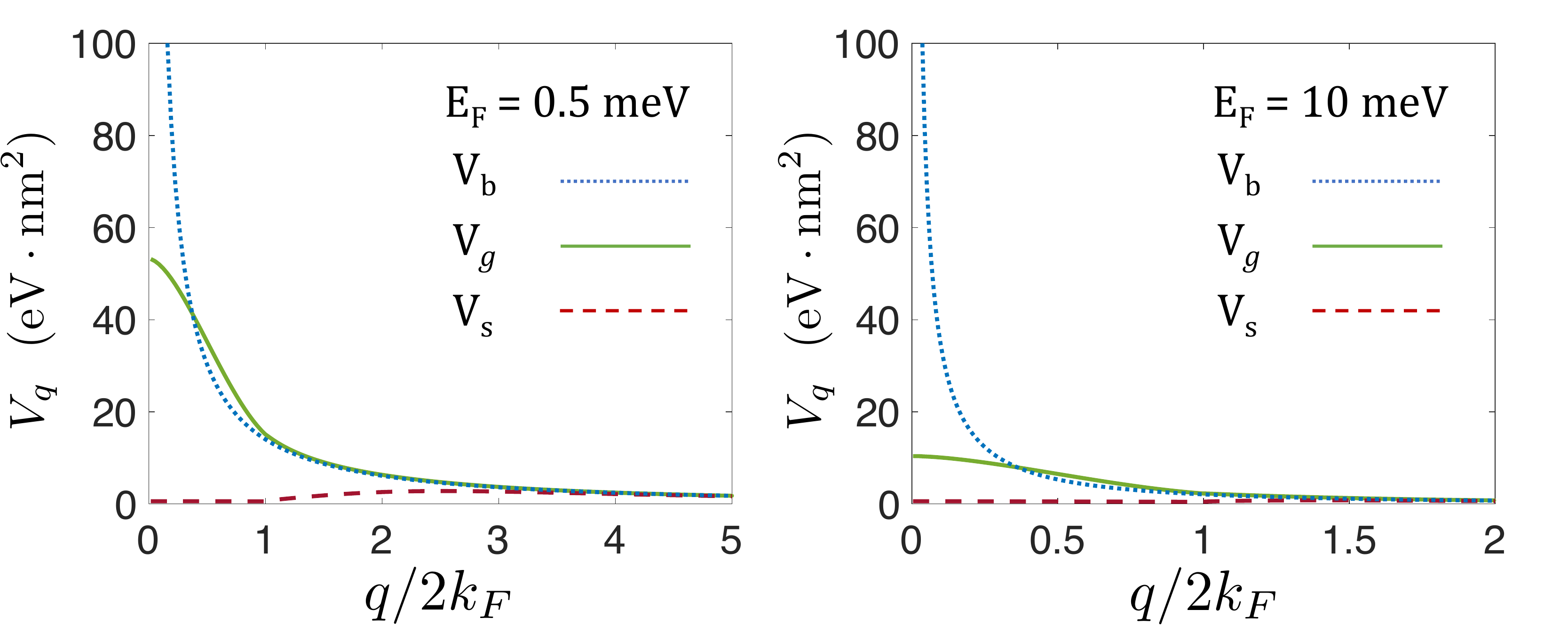}
\caption{Coulomb potentials in 2D systems. The dotted line is the bare Keldysh-Rytova potential ($V_b$), the solid line is partially-screened potential ($V_g$), and the dashed line is the RPA limit of the statically screened potential ($V_s$). } \label{fig:potentials}
\end{figure}

The solid lines in Fig.~\ref{fig:potentials} show results of the partially screened potential, following Eq.~(\ref{eq:vg_numer}) with $q_0 = 0.1$~nm$^{-1}$. The left and right panels correspond to $E_F=0.5$ and 10~meV, respectively. The dashed lines show the static limit of the screened potential when using the RPA in Eq.~(\ref{eq:RPA}), and the dotted line is the Keldysh-Rytova potential  in Eq.~(\ref{eq:KRP}).  As can be seen from the values of the potential at long-wavelengths ($q \rightarrow 0$), the partially screened potential (solid lines) is closer to the bare Keldysh-Rytova potential when $E_F$ is small and to the RPA screened potential when $E_F$ is large.

\section{Correlated trions} \label{sec:tetron_pexciton} 

We analyze  in this section two types of complexes made of trions and CB holes.  The combined trion-hole complex is  referred to as a correlated trion. The calculations we present are based on the SVM-$k$  \cite{l}, where the Keldysh-Rytova potential is used for interactions among particles of the trion, while the partially screened potential in Eq.~(\ref{eq:vg_numer})  is used for interactions that involve the CB hole or  Fermi sea electrons. Later, we will also analyze the changes to the energy of the correlated trion when using the  Keldysh-Rytova potential to describe interactions among all quasiparticles of the composite.

\subsection{ Tetron}  \label{sec:tetron} 
\begin{figure}[]
\includegraphics[width=8.5cm]{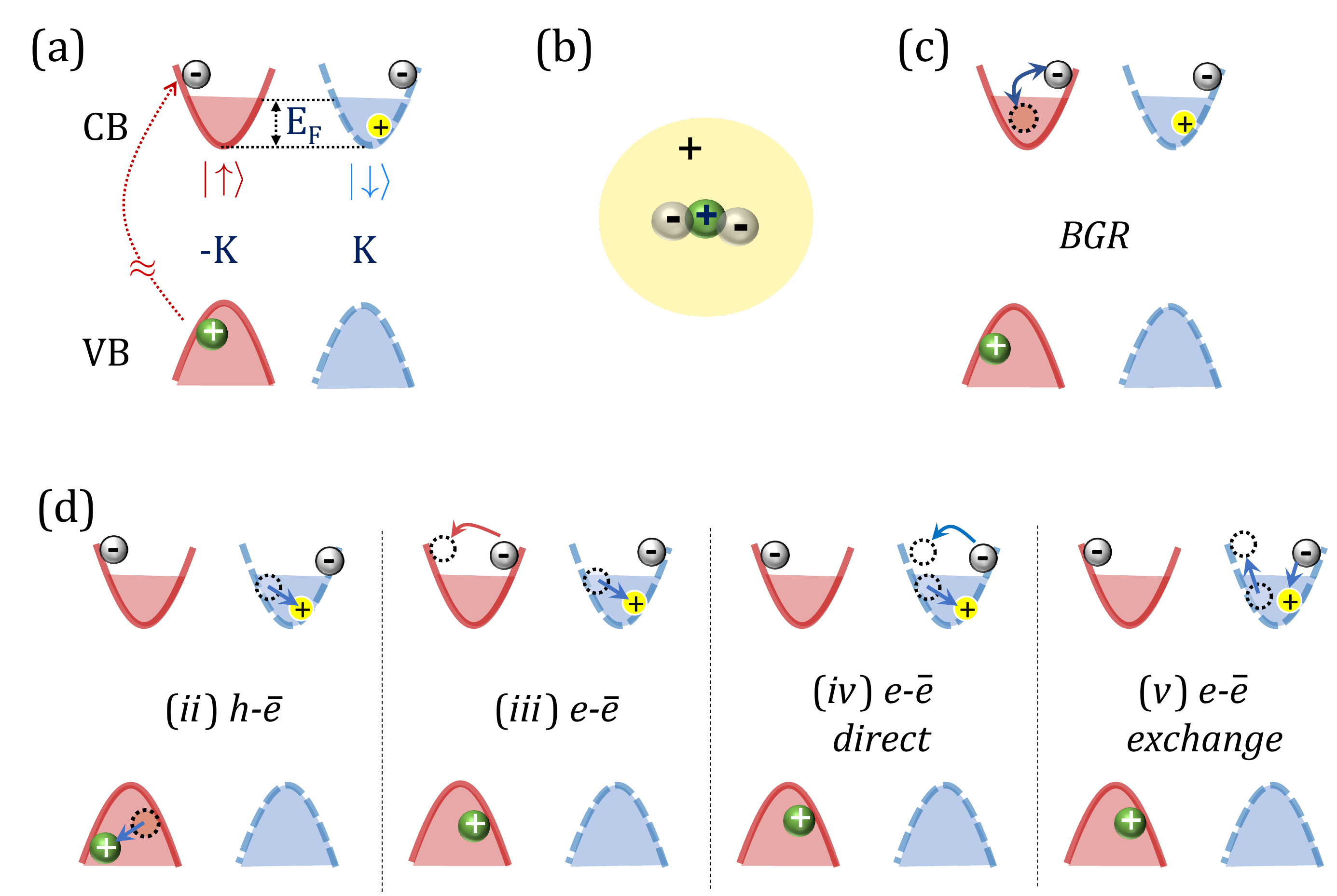}
\caption{(a) Tetron formation following photoexcitation of the valley at $-K$. The photoexcited pair binds to a pulled-out electron from the Fermi sea in $K$ and the left-behind Fermi hole in the CB.  (b) Real-space configuration of the tetron. The CB hole around the core trion is the Pauli-excluded area, depleted of electrons from the Fermi sea in $K$.  (c) Band-gap renormalization due to exchange interaction between the photoexcited electron and Fermi-sea electrons in $-K$. This interaction keeps the electrons in $-K$ away from the complex. (d) Coulomb processes due to interactions with the CB hole at $K$,  denoted by $\bar{e}$. The rightmost two processes are the direct and exchange interaction of the CB hole with the trion's electron that resides in the same valley. The processes in (c) and (d) are considered in addition to the ones in Fig.~\ref{fig:trion}(b).} \label{fig:tetron}
\end{figure}

The first configuration we analyze is the tetron, shown in Figs.~\ref{fig:tetron}(a) and (b).  The spin-up electrons from the CB valley at $-K$ are kept away from the photoexcited electron through exchange interaction. This interaction, shown in Fig.~\ref{fig:tetron}(c), contributes to the self-energy of the photoexcited electron, leading to the celebrated band-gap renormalization (BGR) effect. In addition to the processes we have analyzed in Fig.~\ref{fig:trion}(b), which led to binding of the three particles in the core trion, the tetron includes interactions of these particles with the CB hole, as shown in Fig.~\ref{fig:tetron}(d). The electron-hole exchange process is depicted in the rightmost diagram. Note that the BGR effect of the trion's electron from the valley at $K$ is offset by that of the CB hole (i.e., of the missing electron in the Fermi sea). All in all, the energy of the complex is lowered with respect to the core trion by binding to the CB hole at $K$ and by the BGR of the electron at $-K$.   

\begin{figure}
\includegraphics[width=8.7cm]{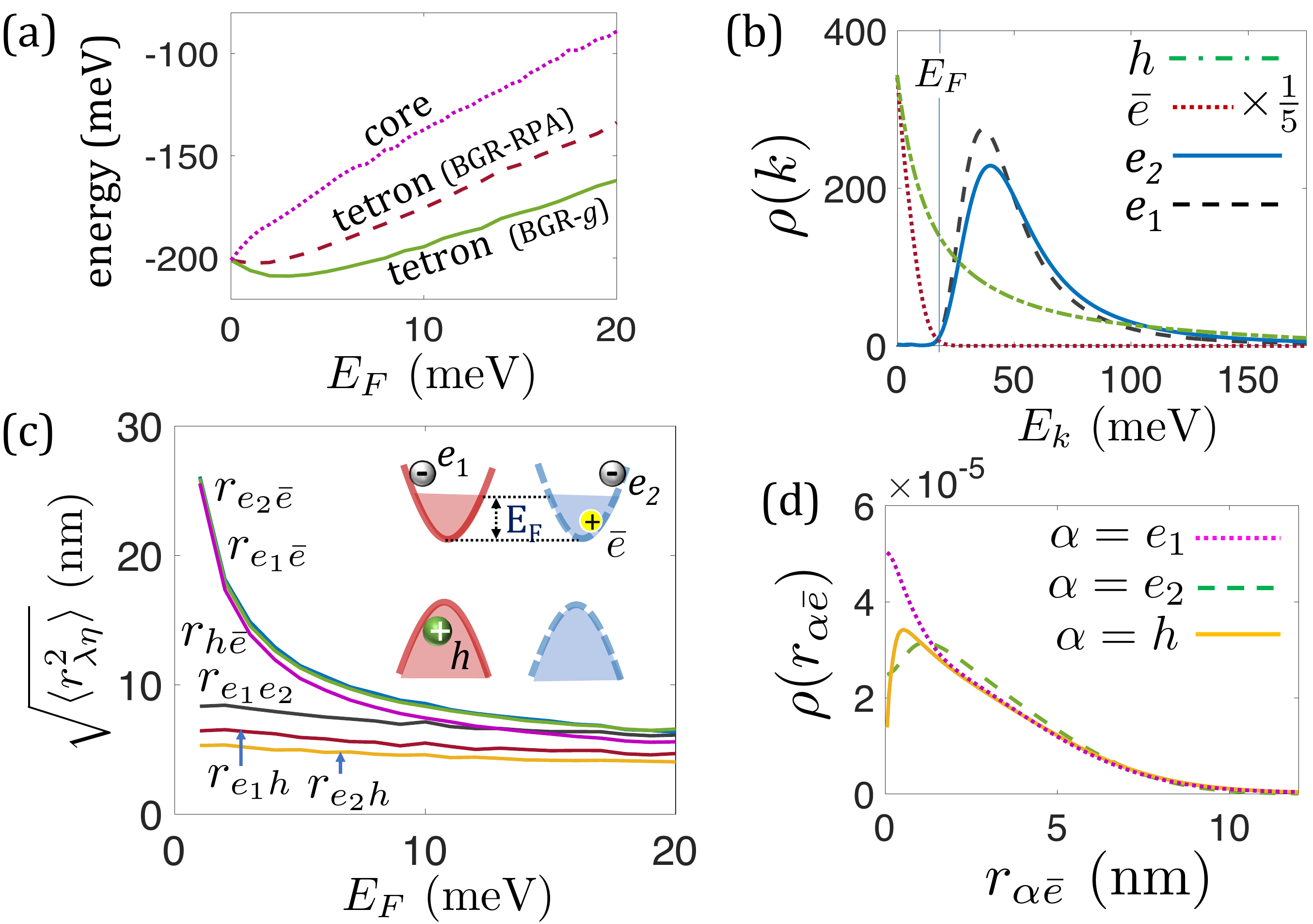}
\caption{(a) Binding energy vs Fermi energy. The dotted line shows the binding energy of the core trion (Fig.~\ref{fig:core}). The dashed and solid lines show the binding energy of the tetron, where the BGR of the photoexcited electron is calculated using the RPA or partially screened potentials, respectively. (b) $k$-space density distributions of particles in the tetron versus kinetic energy when $E_F = 20$~meV. (c) Average inter-particle distances in the tetron. Inset: the tetron with its labeled particles. (d) Density distributions for distances from the CB hole when $E_F = 20$~meV.} \label{fig:tetron_sim}
\end{figure}

Figure~\ref{fig:tetron_sim}(a) shows the binding energy of the tetron, where the  dotted line denotes the binding energy of the non-interacting trion at its core, taken from Fig.~\ref{fig:core}(b). The interactions with the CB hole are handled with the partially-screened potential  in both cases. The  BGR of the photoexcited electron is calculated from \cite{l}
\begin{equation}
\Sigma({\bf k}) = - \sum_{\mathbf{q}} V({\bf q}) f({\bf k - q}),
\label{Eq:BGR}
\end{equation}
where $f({\bf k})$ is the Fermi-Dirac distribution, and $V({\bf q})$ is the Coulomb potential. The solid and dashed lines show the results when using the partially-screened and RPA potentials, respectively. Figure~\ref{fig:tetron_sim}(b) shows the density distributions of each particle in the tetron, extracted by integrating out the wavevector components of all particles but the one of interest from the square-amplitude of the wavefunction \cite{l}. The band filling effect can be seen from the vanishing wavefunction of the  electrons components below the Fermi energy  (solid blue and dashed black lines), whereas that of the CB hole vanishes above the Fermi energy (dotted red line). On the other hand, the VB hole has no $k$-space restriction (dashed-dotted green line). The  distributions of the two electrons are not the same because electron 1 is affected by BRG and electron 2 by the electron-hole exchange with its paired CB hole. 

Figure~\ref{fig:tetron_sim}(c) shows average distances between particles of the tetron, where the BGR is calculated with the partially screened potential.  Increasing the electron density hardly affects the average distances between the particles of the core trion ($r_{e_1e_2}$, $r_{e_1h}$  \& $r_{e_2h}$), from which it is understood that the trion remains intact. The reason for the enhanced binding of the tetron compared with the core trion is the BGR and shrinking size  of the CB hole when the charge density increases (its spatial extent is commensurate with $1/k_F$). This behavior is evidenced by the decreasing distance between the CB hole and the trion particles ($r_{e_1\bar{e}}$, $r_{e_2\bar{e}}$  \& $r_{h\bar{e}}$).  Figure~\ref{fig:tetron_sim}(d) shows density distributions for distances from the CB hole when $E_F = 20$~meV. The dip at $r=0$ between the CB and VB holes is due to their Coulomb repulsion (solid line), whereas the dip at $r=0$ between the CB hole and its paired electron is due to effect of the repulsive electron-hole exchange interaction (dashed line) \cite{l}. As shown by the rightmost diagram of Fig.~\ref{fig:tetron}(d), the \textit{e-}$\bar{e}$ exchange interaction describes a process in which a pair is annihilated and a new one is generated. 
A detailed calculation in real space shows that the amplitude of this scattering is proportional to the squared amplitude of the wavefunction at short distances. Thus,  the annihilation probability is larger when the distance between the electron and its CB hole is small, leading to the dip at $r=0$.

\begin{figure}
\includegraphics[width=8.7cm]{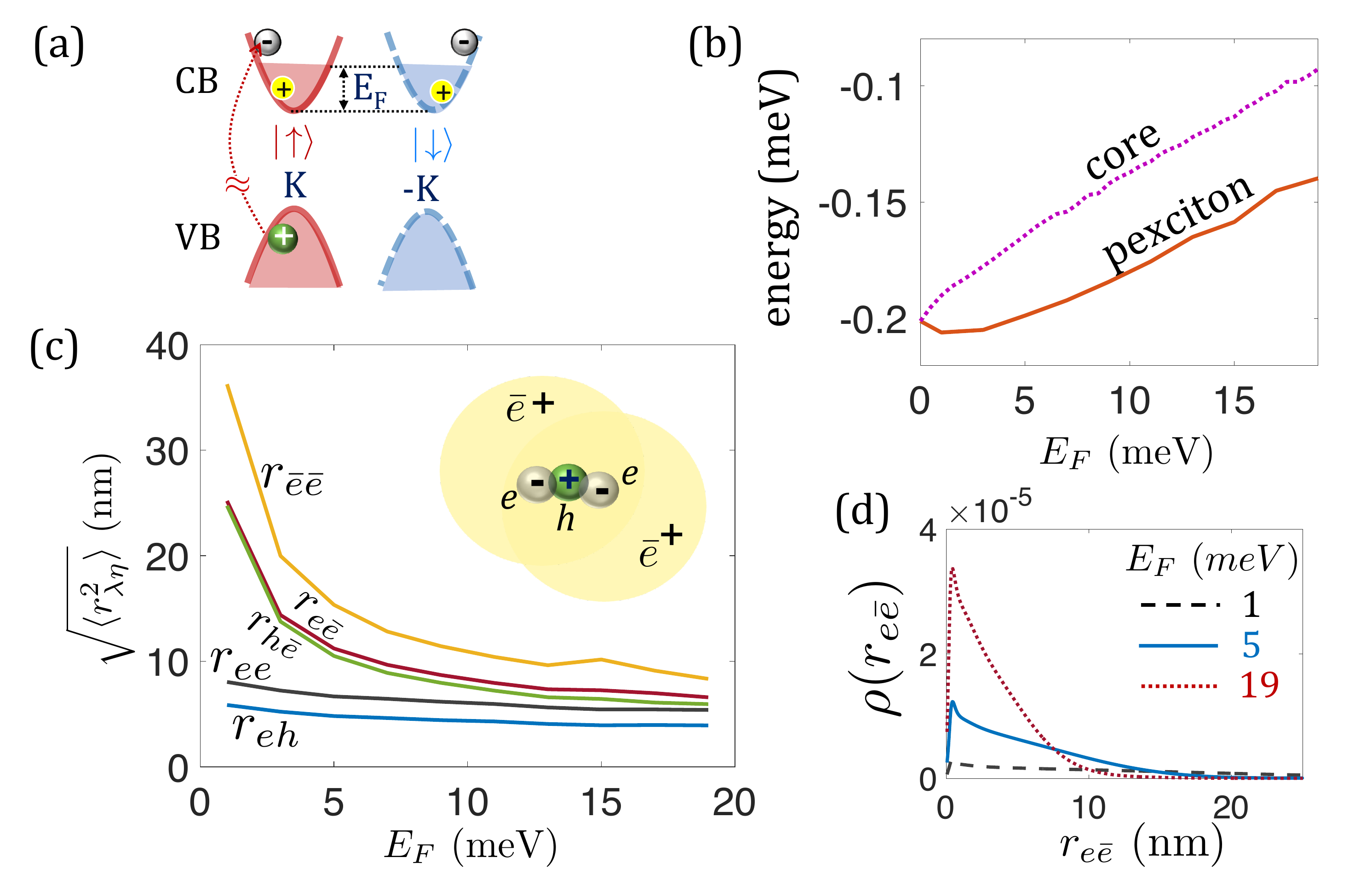}
\caption{(a) The five-body correlated trion. (b) Binding energy vs Fermi energy, where the solid (dotted) line shows the binding energy of the correlated (core) trion. That is, with and without the CB holes. (c) Average inter-particle distances in the complex. Inset: Labels of particles in the complex (shown in real space). (d) The density distribution for the distance between an electron and its paired CB hole.} \label{fig:5_sim}
\end{figure}

\subsection{Pexciton }
Our next composite is the five-body complex (``pexciton'') in Fig.~\ref{fig:5_sim}(a), comprising a core trion and two CB holes. This is an alternative view for the tetron, wherein instead of BGR of the photoexcited electron, we pair this electron with a CB hole.  For example, we assume that the photoexcitation involves a shakeup process that drives away electrons with the same spin and valley of the photoexcited electron, leading to creation of the CB hole in the $-K$ valley. The electron from the time-reversed valley in $K$ is pulled out of the Fermi sea, resulting in the  second CB hole. 

Figure~\ref{fig:5_sim}(b) shows the binding energy of the pexciton (solid line), where the  dotted line shows the binding energy of the non-interacting trion at its core, taken from Fig.~\ref{fig:core}(b). Coulomb interactions with CB holes are handled with the partially-screened potential. Figure~\ref{fig:tetron_sim}(c) shows average distances between particles of the complex. Here, the two \textit{e-}$\bar{e}$ pairs are equivalent because we do not consider the \textit{e-h} exchange interaction of the photoexcited pair (we only consider the strong exchange interaction of the pair \textit{e-}$\bar{e}$).  As before, increasing the electron density hardly affects the average distances between the particles of the core trion ($r_{ee}$ \& $r_{eh}$), from which it is understood that the trion remains intact. The reason for the enhanced binding of the pexciton compared with the core trion is the shrinking size  of its CB holes when the charge density increases. The shrinkage is shown in Fig.~\ref{fig:5_sim}(d), where we plot the density distribution of the electron-hole CB pair ($r_{e\bar{e}}$) for three densities. As mentioned before, the dip at $r_{e\bar{e}}=0$ is due to repulsive electron-hole exchange interaction  of the pair \textit{e-}$\bar{e}$  \cite{l}. 

\subsection{Other potential forms }
The results shown in Figs.~\ref{fig:tetron_sim} and \ref{fig:5_sim} were calculated by assuming that interactions with CB holes are handled with the partially screened potential and interactions  among the two CB electrons and VB hole of the core trion  with the bare potential. As was shown in Fig.~\ref{fig:potentials}, the difference between the bare and partially-screened potentials becomes more evident at large densities due to larger weight of screening. The difference between the potentials leads to smaller binding energy of tetrons or pexcitons at large densities, as shown by Figs.~\ref{fig:tetron_sim}(a) and \ref{fig:5_sim}(b). The loss of binding energy from $k$-space restriction of the electrons in the core trion cannot be compensated by increased binding to the CB hole because the latter is calculated with a weaker potential. 

\begin{figure}
\includegraphics[width=8.7cm]{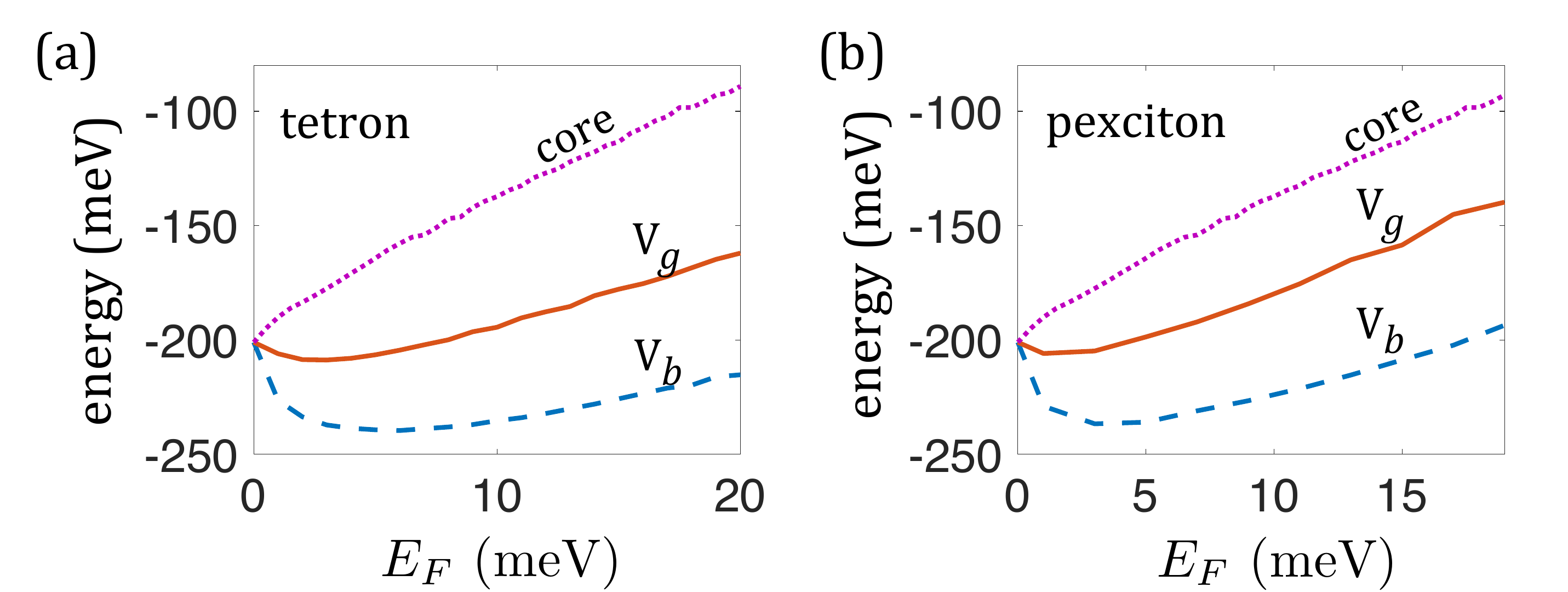}
\caption{Binding energies of correlated trions vs Fermi energy. (a) and (b) show the tetron and pexciton cases. The dotted lines show the binding energy of the core trion (Fig.~\ref{fig:core}). The solid (dashed) lines show results when  interactions with CB holes are handled with the partially-screened (bare) potential.} \label{fig:e45}
\end{figure}

To check the results with a different potential choice, we repeat the calculations but  assume that interactions with CB holes are also handled with the bare Keldysh-Rytova potential. The dashed lines in Figs.~\ref{fig:e45}(a) and (b) show the calculated binding energies for tetrons and pexcitons, respectively.  We notice a fast increase in binding energy when the Fermi energy increases from zero to a few meV, followed by a moderate decrease in binding energy when the Fermi energy continues to increase. In comparison, the results with the partially-screened potential show the opposite trend (solid lines). Namely, a \textit{moderate} increase at small Fermi energies  vs a \textit{fast} decrease at large Fermi energies.

The fast initial increase in binding energy when using the bare potential is caused by strong binding to the CB hole. The latter is restricted to states with very small wavevectors at vanishing densities, and as such, it is mostly influenced by the  long-wavelength regime in which the Keldysh-Rytova potential diverges ($q \rightarrow 0$).  The use of the partially-screened potential mitigates this effect.  On the other hand, the milder change in binding energy at larger Fermi energy when using the bare potential is caused by better offset between the effects from decreased $k$-space of electrons and increased $k$-space of CB holes. 

Experiments at zero magnetic field show that the change in binding energy of trions, seen through energy shift of their optical transitions,  is moderate throughout the range of Fermi energies we study here (corresponding to charge densities between 0 and 4$\times$10$^{12}$~cm$^{-2}$) \cite{Wang_NanoLett17,Smolenski_PRL19,Wang_PRX20,Liu_PRL20,Liu_NatComm21,Li_NanoLett22}. Within the context of the simulation results, such behavior implies that interactions with CB holes are overestimated when modeled by the  bare potential at small charge densities, but they are reasonably correct at larger densities. At first, this result seems counterintuitive because screening should play larger role at larger densities. However, Figs.~\ref{fig:tetron_sim}(c)  and \ref{fig:5_sim}(c) also reveal that the average distances of CB holes from other particles in the complex become small and even comparable to the size of the core trion at large densities. The screening effect at such short distances should indeed be mitigated. Future studies are needed to better describe the Coulomb interactions of composite excitonic states in doped semiconductors. While we are still not out of the woods yet, we can say that the screening effect is far weaker than the one modeled by RPA \cite{Scharf_JPCM19,VanTuan_PRX17}. 

\section{Hexciton } \label{sec:hex} 

We have and will continue to consider excitonic states with one VB hole, applicable in cases that the electron density is much larger than the density of photoexcited \textit{e-h} pairs.  So far, we have modeled complexes with two CB electrons, and our next step is to study complexes with three CB electrons. As we will explain, this scenario is applicable in  electron-rich WSe$_2$, whose  unique optical transitions were first observed by Jones  \textit{et al.}  \cite{Jones_NatNano13}. The possibility of having a complex with three CB electrons and one VB hole was commented by Barbone \textit{et al.}, who suggested that the next charging state of the trion can explain the photoluminescence of electron-rich WSe$_2$  ~\cite{Barbone_NatComm18}.  As we will show, this interpretation is correct when augmented with the understanding that the binding between a VB hole and three distinguishable electrons is facilitated by CB holes, which provide a positive charge background. 

\begin{figure}[]
\includegraphics[width=8.7cm]{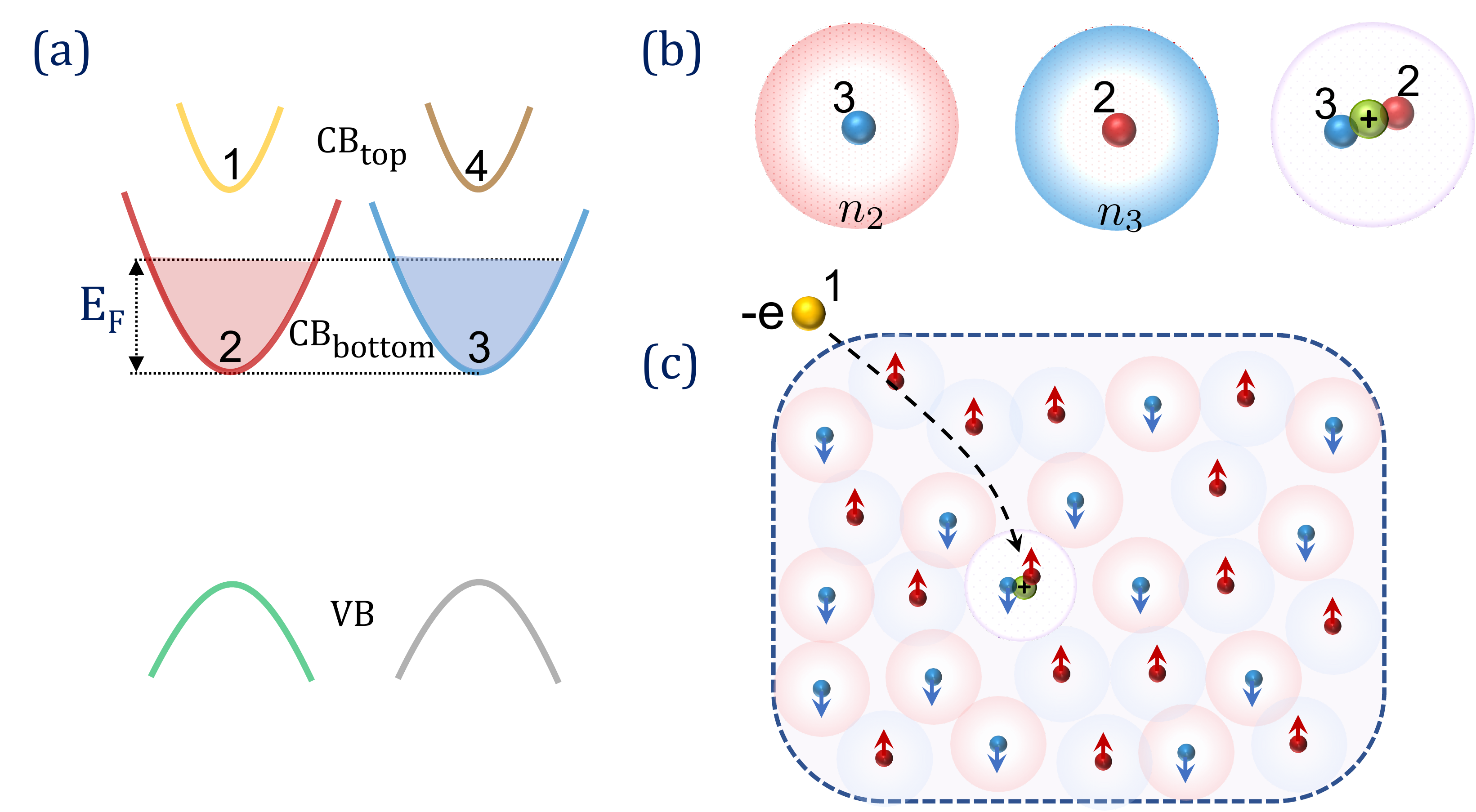}
\caption{(a) Band diagram of WSe$_2$ monolayers, showing four spin-polarized CB valleys, where the bottom two are populated (reservoirs 2 and 3).  (b) Diagrams of the charge distribution around an electron or a trion.  The charge at the center is $-e$, surrounded by an electron-depleted region due to Coulomb repulsion. The Pauli exclusion principle means that it is more probable to find electrons from reservoir 2 around an electron from reservoir 3 (and vice versa). Since the trion comprises electrons from both reservoirs, its surrounding is more positively charged compared with the surroundings of electrons from the Fermi gas. (c) A trion in a gas of electrons from reservoirs 2 and 3.  The `positive' charge bubble around the trion is an energetically favorable place for an electron from valleys 1 or 4.} \label{fig:Hscheme}
\end{figure}

We begin by considering the energy band diagram in Fig.~\ref{fig:Hscheme}(a). This diagram resembles that of WSe$_2$ monolayers, wherein the CB has four spin-polarized valleys, where the bottom two are populated (reservoirs 2 and 3) and the top two are optically active (valleys 1 and 4). Figure~\ref{fig:Hscheme}(b) shows charge distribution schemes around Fermi-sea electrons and an optically-dark trion whose electrons are from reservoirs 2 and 3.  The density of the electron cloud around an electron from the sea or the dark trion is proportional to
\begin{eqnarray}
\rho_e(r) &\propto& - \tfrac{1}{2}[ g_i(r) + g_d(r)] n, \nonumber \\
\rho_T(r) &\propto& - g_i(r) n,   \label{eq:e_g}
\end{eqnarray}
respectively. $n$ is the electron density, $g_i(r)$ is the pair distribution function for indistinguishable electrons (i.e.,  from the same reservoir), and $g_d(r)$ is the pair distribution function for distinguishable electrons from different reservoirs. When having two equal reservoirs with spin-up and spin-down electrons, $g_i(r) = g_{\uparrow\uparrow}(r) = g_{\downarrow\downarrow}(r)$ and $g_d(r) = g_{\uparrow\downarrow}(r) = g_{\downarrow\uparrow}(r)$. The Pauli exclusion  principle mandates that $g_d(r)>g_i(r)$. Namely, the joint probability of finding two particles at distance $r$ is larger if the electrons are distinguishable.

Let us now assume that an electron from the unpopulated reservoir 1 is added to a system made of the dark trion and electron gas from reservoirs 2 and 3, as shown in Fig.~\ref{fig:Hscheme}(c).  Whereas the charge of a bare trion is the same as that of an electron from the Fermi sea ($-e$), the electron depletion in the region next to the trion is stronger due to the Pauli exclusion principle (electrons from reservoirs 2 and 3 are excluded). Choosing between being near an electron from the Fermi gas or the trion, the added electron from reservoir 1 prefers the trion because it is surrounded by a  relatively positive charge bubble in an otherwise homogenous electron gas. The result is then a composite complex comprising the correlated trion and satellite (distinguishable) electron.  

In absorption-type experiments, one can probe the resonance of the hexciton if the added electron is the photoexcited electron \cite{h}.  This condition is met in WSe$_2$ monolayers, wherein the photoexcited electron is from the top CB valley. Furthermore,  the effective mass of the photoexcited electron is about 30\% less than the electron effective mass  in the bottom CB valley \cite{Kormanyos_2DMater15}.  Thus, the trion at the core of the composite is made of the photoexcited VB hole and two heavier  electrons from the bottom CB valleys. 

\begin{figure*}
\includegraphics[width=17cm]{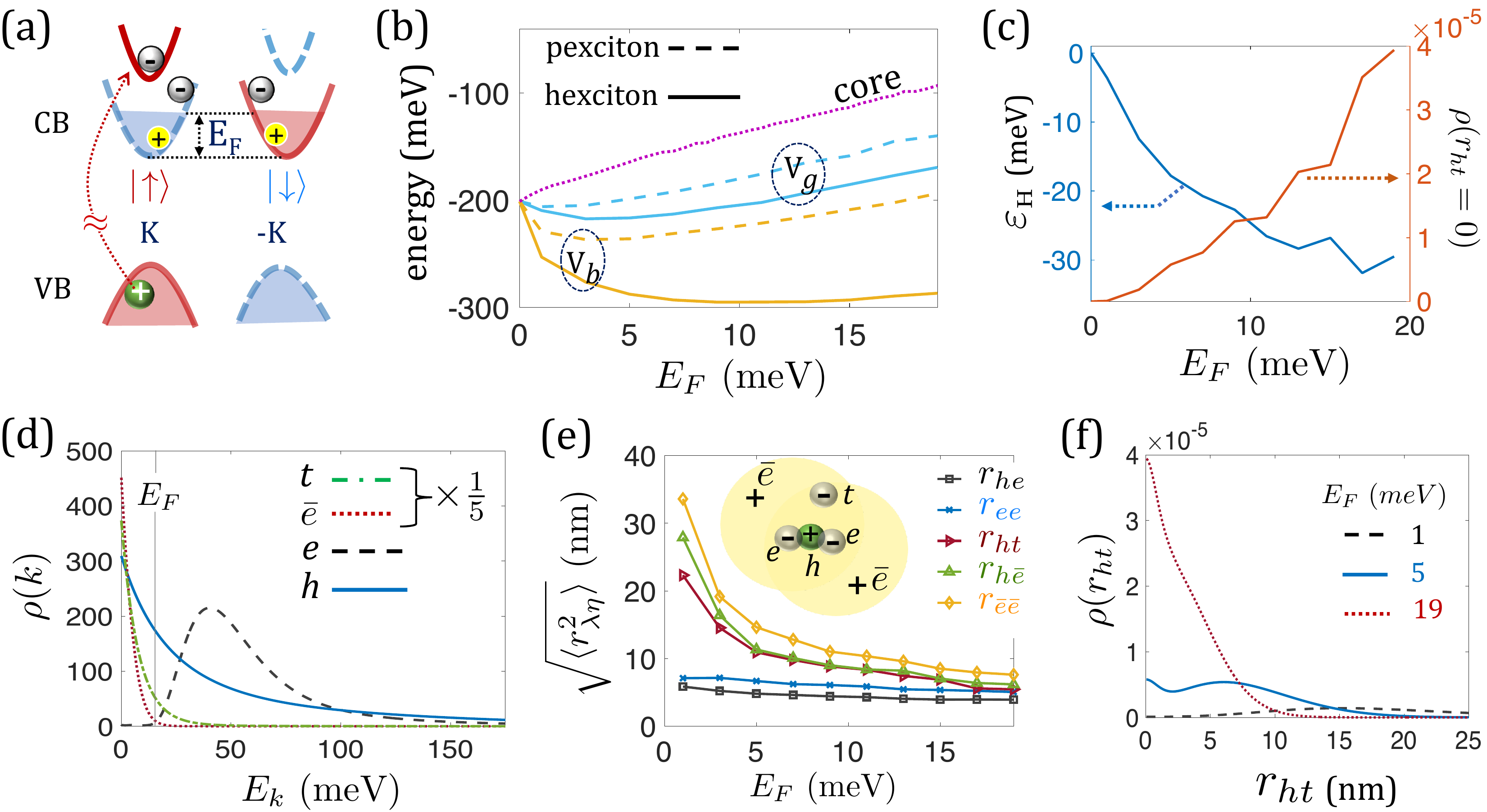}
\caption{(a) The hexciton composite, made of the photoexcited pair and two more CB electron-hole pairs. (b) Binding energy vs Fermi energy, where the solid lines show the binding energy of the hexciton calculated with the bare and screened potentials. The dashed lines show the respective binding energies of the pexciton (i.e., without the satellite electron), and the dotted line shows the binding energy of the core trion. Continuing with simulation results of the screened potential, (c) Left: Energy difference between the hexciton and pexciton,  representing the  binding energy of the satellite electron. Right: Overlap between VB hole and satellite electron. (d) $k$-space density distributions of particles in the hexciton versus their kinetic energy when $E_F = 19$~meV. (e) Average inter-particle distances in the complex. Inset: Labels of particles in the hexciton. (f) Density distribution of the distance between the satellite electron and VB hole} \label{fig:6_sim}
\end{figure*}

Figure~\ref{fig:6_sim}(a) shows the hexciton composite we are about to analyze  using the SVM-\textit{k} \cite{l}. It comprises  the photoexcited \textit{e-h} pair and two more \text{e-}$\bar{e}$ pairs. The solid lines in Fig.~\ref{fig:6_sim}(b) show the energy of the composite, calculated when interactions with CB holes and/or the satellite electron are handled by the bare or screened potentials. The dashed lines show the respective binding energies of the pexciton, which is essentially a hexciton ionized of its satellite electron. As before, the dotted line is used as a reference to show the binding energy of the core (dark) trion. Continuing with the screened potential simulation results, the energy difference between the hexciton and pexciton is shown in Fig.~\ref{fig:6_sim}(c) along with the overlap between VB hole and satellite electron. This overlap, a measure of the oscillator strength of the hexciton's optical transition, is commensurate with the electron density because the positively charge cloud around the trion shrinks (CB holes) and leads to tighter  binding of the satellite electron to the complex. The observed behavior is then similar to the one seen in experiment \cite{h}. The redshift of the optical transition when increasing the charge density signifies the added binding energy due to tighter binding of the satellite electron.  The amplification of the optical transition when increasing the charge density signifies the stronger overlap. 

Figure~\ref{fig:6_sim}(d) shows the density distributions when $E_F = 19$~meV. The restricted space of the electrons (\textit{e}) and  CB holes ($\bar{e}$)  is evidenced from the vanished distribution of the electron below the Fermi energy (dashed black line), and the vanished distribution of its CB hole above the Fermi energy (dotted red line). On the other hand, the VB hole and top-valley (satellite) electron distribute across the $k$-space.  The wider distribution of the VB hole is a measure of its much larger contribution to the binding energy compared with the satellite electron. The average inter-particle distances are shown in Fig.~\ref{fig:6_sim}(e), where the inset shows the labeled particles of the hexciton in real space.  Finally, Fig.~\ref{fig:6_sim}(f) shows the density distribution for distances between the VB hole and satellite electron at three charge densities. These distributions are an alternative view to see the tighter binding of the satellite electron to the hexciton when the charge density increases. 

\subsection*{A simplified model for binding of the satellite electron}
The SVM-$k$ is a useful tool to study the interaction of a photoexcited \textit{e-h} pair with the Fermi sea(s). However, the calculation is time consuming even for a system with  few quasiparticles. For instance, it takes several weeks to calculate the hexciton state at a certain Fermi energy using a single high-end processor. We describe a simplified one-body model to mitigate this problem, from which we study the binding energy and charge distribution of the satellite electron. The calculation takes a few seconds and can be readily implemented, but still, it quantitatively captures the behavior of the satellite electron.  
 
The binding energy and charge distribution of the satellite electron are studied by considering the pexciton as an energy reference level, and by replacing its five quasiparticles with a fixed charge distribution that represents the core trion and two CB holes. The charge distribution of the core trion ($-e$) is modeled by
\begin{equation}
\rho_T(r) = -\frac{|e|}{\pi a_T^2} e^{-r/a_T}\,,
\end{equation}
where $a_T$ is the effective radius of the core trion. Using the relatively fast SVM-$k$ calculation of the bare trion, one can extract the value $a_T$ ($\sim$2~nm).  

Charge distribution of the two CB holes is evaluated by following the discussion of Eq.~(\ref{eq:e_g}), where $ \rho_h(r) \propto \rho_e(r)-\rho_T(r)$.  Using Eq.~(\ref{eq:g}) and requiring that the charge is $+2e$, we get that  
\begin{equation}
 \rho_h(r) =  2\frac{|e|}{\pi} \frac{J_1^2(k_F r)}{ r^2}.
\end{equation}
The effective potential seen by the satellite electron due to the `static' pexciton charge distribution is
\begin{equation}
V^*({\bf r}) =   \int d{{\bf r}'} V({\bf r - r}') \left( \rho_T({\bf r}') + \rho_h({\bf r}') \right) /|e| \,,
\end{equation}  
yielding the simple formula for the potential in momentum space,
\begin{equation} 
 V^*( {\bf q})=  \frac{A}{|e|}   V({\bf q})  \left(\rho_T({\bf q}) + \rho_h({\bf q})   \right).
\end{equation}  
$V({\bf q})$  can be the bare or partially-screened Coulomb potential, discussed in Sec.~\ref{sec:screening}. $\rho_T({\bf q})$ and $\rho_h({\bf q})$ are Fourier transforms of the effective charge distributions of the trion and CB hole,
\begin{eqnarray}
\rho_T({\bf q}) &=& - \frac{2|e|}{A \left( 1+ a_T^2 q^2 \right)^{3/2}}, \nonumber \\
 \rho_h({\bf q}) &=&   \frac{4|e|}{A \pi }  \Theta (2k_F -q) \times \nonumber \\ && \left(  \arccos \left(\frac{q}{2 k_F}\right)- \frac{q}{2 k_F} \sqrt{1-\frac{q^2}{4k_F^2}} \right).
\end{eqnarray}

\begin{figure}
\centering
\includegraphics[width=8cm]{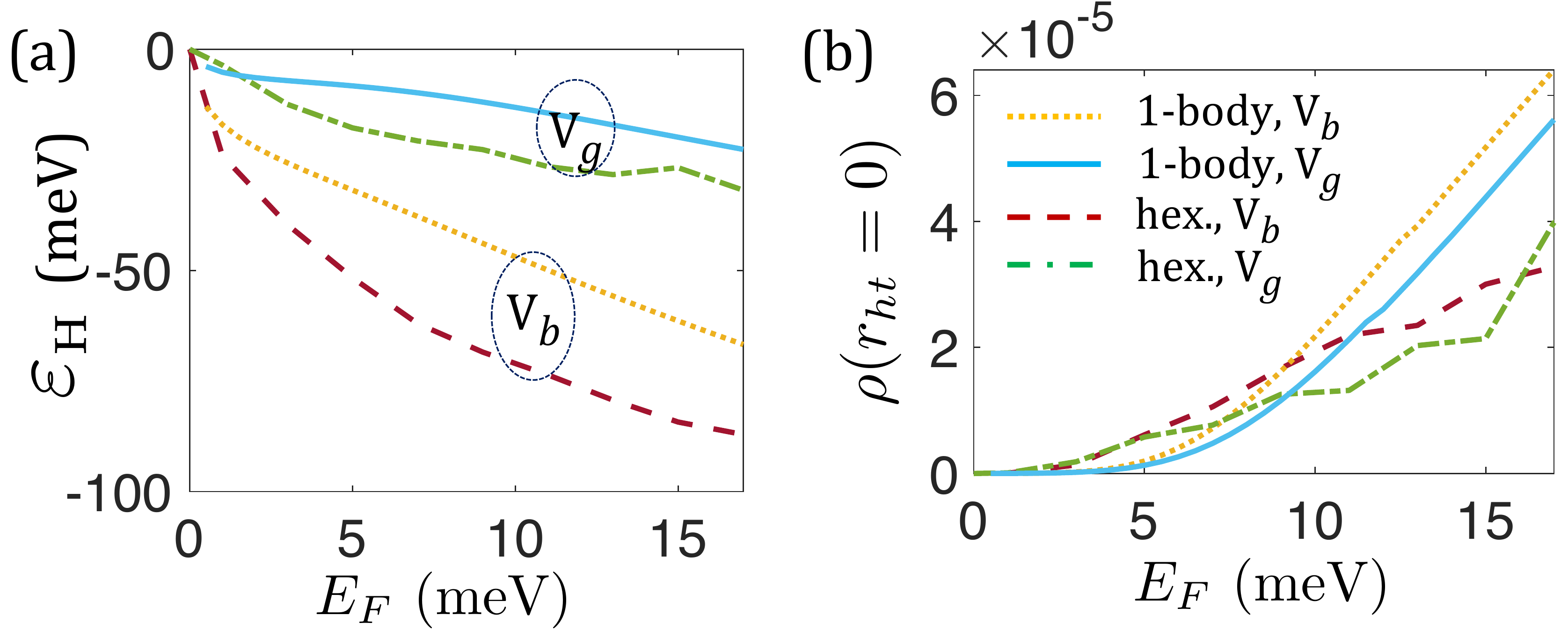}
\caption{ (a) Binding energy of the satellite electron versus Fermi energy. Results are shown for calculations with the bare and partially-screened potentials. The smooth lines show results of the effective one-body model (solid and dotted lines), whereas the other two show results from the energy difference between hexciton and pexciton (dashed and dashed-dotted). (b) The probability to find the satellite electron at the origin in the 1-body effective model, or its overlap with the VB hole obtained from the hexciton SVM-k calculation. }\label{fig:OnePartMod} 
\end{figure}

Figure~\ref{fig:OnePartMod}(a) shows the obtained binding energies calculated with the bare and partially-screened potentials. The smooth lines (solid and dotted) show results of the effective one-body model, and the other lines show the energy difference between the hexciton and pexciton (dashed and dashed-dotted lines). The calculated binding energies from the one-body model are smaller because treating the CB holes as rigid bodies is less effective in lowering the energy of the system. The motions of the CB holes in the pexciton help to screen the repulsive interaction between the core trion and satellite electron. This effect also explains why the overlap factors are somewhat smaller when calculated through the hexciton state, as shown in Fig.~\ref{fig:OnePartMod}(b).  Nonetheless, the  simplified one-body model captures the behavior of the satellite electron reasonably well, and doing so with orders of magnitude shorter computational time.

\section{Conclusions and Outlook}  \label{sec:conc}

We have considered several factors affecting the optical properties of electron-doped transition-metal dichalcogenides monolayers with several distinguishable energy pockets in the conduction band. We have discussed the effects of Pauli blocking, disorder, reduced screening around the composite excitonic state, as well as the exchange interaction between an electron and its paired CB hole. While the theory captures important experimental observations, the calculated results are still not in exact agreement when it comes to the variation of the trion binding energy when charge is added to the monolayer. The reasons for this mismatch can come from several factors which we plan to investigate in the future.  These include better modeling of the screening effect and of disorder at small electron densities. For example, the screening effect can be accounted for by using the bare Coulomb potential but by treating the charge of a CB hole as a density-dependent parameter. In addition, the  SVM-$k$ model can be further improved by introducing scattering states that can couple between the trion and exciton states \cite{Chang_PRB19}. 

\acknowledgments{This work was supported by the Department of Energy, Basic Energy Sciences, Division of Materials Sciences and Engineering under Award DE-SC0014349 (DVT), and by the Office of Naval Research under Award N000142112448 (HD).}

\appendix
\section{Compiled list of parameters}\label{app:params}

 Throughout this work and in Refs.~\cite{h,s}, we have used parameters of monolayer WSe$_2$ embedded in hexagonal-boron nitride.  
 \begin{enumerate}
 \item  The dielectric constant of hexagonal-boron nitride is $\epsilon_b=3.8$.
 
 \item The polarizability of the monolayer is $r_0=1.18$~nm ($r^{\ast}=\epsilon_b r_0 =4.5$~nm). 
 
 \item The effective masses are 0.36$m_0$ for VB holes and 0.4$m_0$ ($0.29m_0$) for electrons in the bottommost (top) CB valleys \cite{Kormanyos_2DMater15}.  
 
 \item We have used $q_0=0.1$~nm$^{-1}$ when modeling the screened potential. The Fourier-Bessel expansion of the screened potential includes 30 terms and its cutoff wavenumber is $q_c = 5$~nm$^{-1}$ (see Ref.~\cite{l} for details).

 \item  The results of the one-body model in Fig.~\ref{fig:OnePartMod} were calculated by using $a_T=2$~nm for the radius of the trion, extracted from the behavior of bare trion at $E_F=0$.
 
 \item 800 basis functions are used in the SVM-$k$  to calculate tetrons, and 1500 to calculate the pexcitons or hexcitons.

 \end{enumerate}

The conclusions we have made remain qualitatively similar if we were to use the  polarizability and effective mass parameters of other monolayers. Yet, it is important to keep in mind that the satellite electron in the hexciton is the one with smaller mass.

\end{document}